\documentclass[french,english,aps,manuscript,preprintnumbers]{revtex4-1}
\usepackage[T1]{fontenc}
\usepackage[latin9]{inputenc}
\usepackage{float}
\usepackage{amssymb}
\usepackage{graphicx}
\usepackage{esint}

\makeatletter

\providecommand{\tabularnewline}{\\}

\@ifundefined{showcaptionsetup}{}{%
 \PassOptionsToPackage{caption=false}{subfig}}
\usepackage{subfig}
\makeatother

\usepackage{babel}
\makeatletter
\addto\extrasfrench{%
   \providecommand{\fg}{\ifdim\lastskip>\z@\unskip\fi~\frqq}%
}

\makeatother
\begin{document}

\title{Effects of a minimal length on the thermal properties of a Dirac
oscillator}

\author{Abdelmalek Boumali}

\email{boumali.abdelmalek@gmail.com}

\selectlanguage{english}%

\affiliation{Laboratoire de Physique Appliquée et Théorique, \\
 Université de Tébessa, 12000, W. Tébessa, Algeria.}

\author{Lyazid Chetouani}

\email{lyazidchetouani@gmail.com}

\selectlanguage{english}%

\affiliation{Département de Physique, Université de Constantine, W. Constantine,
25000, Algeria.}

\author{Hassan Hassanabadi}

\email{h.hasanabadi@shahroodut.ac.ir}

\selectlanguage{english}%

\affiliation{Department of Physics, University of Shahrood, Shahrood Iran.}

\date{\today}
\begin{abstract}
The effect of the minimal length on the thermal properties of a Dirac
oscillator is considered. The canonical partition function is well
determined by using the method based on the Epstein Zeta function.
Through this function, all thermodynamics properties, such as the
free energy, the total energy, the entropy, and the specific heat,
have been determined. Moreover, this study leads to a minimal length
in the interval $10^{-16}<\triangle x<10^{-14}\mbox{m}$ with the
following physically acceptable condition $\beta>\beta_{0}=\frac{1}{m_{0}^{2}c^{2}}$.
We show that this condition is obtained directly through the properties
of the Epstein Zeta function, and the minimal length $\Delta x$ coincide
with the reduced Compton wavelength $\bar{\lambda}=\frac{\hbar}{m_{0}c}$.
\end{abstract}



\maketitle

\section{Introduction}

The Dirac relativistic oscillator is an important potential both for
theory and application. It was for the first time studied by Ito et
al\cite{1}. They considered a Dirac equation in which the momentum
$\vec{p}$ is replaced by $\vec{p}-im\beta\omega\vec{r}$, with $\vec{r}$
being the position vector, $m$ the mass of particle, and $\omega$
the frequency of the oscillator. The interest in the problem was revived
by Moshinsky and Szczepaniak \cite{2}, who gave it the name of Dirac
oscillator (DO) because, in the non-relativistic limit, it becomes
a harmonic oscillator with a very strong spin-orbit coupling term.
Physically, it can be shown that the (DO) interaction is a physical
system, which can be interpreted as the interaction of the anomalous
magnetic moment with a linear electric field \cite{3,4}. The electromagnetic
potential associated with the DO has been found by Benitez et al \cite{5}.
The Dirac oscillator has attracted a lot of interest both because
it provides one of the examples of the Dirac's equation exact solvability
and because of its numerous physical applications (see \cite{6} and
references therein). Recently, Franco-Villafane et al \cite{7} exposed
the proposal of the first experimental microwave realization of the
one-dimensional DO. Quimbay et al \cite{8,9} show that the Dirac
oscillator can describe a naturally occurring physical system. Specifically,
the case of a two-dimensional Dirac oscillator can be used to describe
the dynamics of the charge carriers in graphene, and hence its electronic
properties. This idea has been also proved in the calculations of
the thermal properties of graphene using Zeta function \cite{10}.

The unification between the general theory of relativity and the quantum
mechanics is one of the most important problems in theoretical physics.
This unification predicts the existence of a minimal measurable length
on the order of the Planck length. All approaches of quantum gravity
show the idea that near the Planck scale, the standard Heisenberg
uncertainty principle should be reformulated. The minimal length uncertainty
relation has appeared in the context of the string theory, where it
is a consequence of the fact that the string cannot probe distances
smaller than the string scale $\hbar\sqrt{\beta}$, where $\beta$
is a small positive parameter called the deformation parameter. This
minimal length can be introduced as an additional uncertainty in position
measurement, so that the usual canonical commutation relation between
position and momentum operators becomes$\left[\hat{x},\hat{p}\right]=i\hbar\left(1+\beta p^{2}\right)$.
This commutation relation leads to the standard Heisenberg uncertainty
relation $\triangle\hat{x}\triangle\hat{p}\geq i\hbar\left(1+\beta\left(\triangle p\right)^{2}\right)$,
which clearly implies the existence of a non-zero minimal length $\triangle x_{\mbox{min}}=\hbar\sqrt{\beta}$.
This modification of the uncertainty relation is usually termed the
generalized uncertainty principle (GUP)or the minimal length uncertainty
principle\cite{11,12,13,14}. Investigating the influence of the minimal
length assumption on the energy spectrum of quantum systems has become
an interesting issue primarily for two reasons. First, this may help
to set some upper bounds on the value of the minimal length. In this
context, we can cite some studies of the hydrogen atom and a two dimensional
Dirac equation in an external magnetic field. Moreover, the classical
limit has also provided some interesting insights into some cosmological
problems. Second, it has been argued that quantum mechanics with a
minimal length may also be useful to describe non-point-like particles,
such as quasi-particles and various collective excitations in solids,
or composite particles (see Ref \cite{15} and references therein).
Nowadays, the reconsideration of the relativistic quantum mechanics
in the presence of a minimal measurable length have been studied extensively.
In this context, many papers were published where a different quantum
system in space with Heisenberg algebra was studied. They are: the
Abelian Higgs model \cite{16}, the thermostatics with minimal length
\cite{17}, the one-dimensional Hydrogen atom \cite{18}, the casimir
effect in minimal length theories \cite{19}, the effect of minimal
lengths on electron magnetism \cite{20}, the Dirac oscillator in
one and three dimensions \cite{21,22,23,24,25}, the solutions of
a two-dimensional Dirac equation in presence of an external magnetic
field \cite{26}, the noncommutative phase space Schrödinger equation\cite{27},
Schrödinger equation with Harmonic potential in the presence of a
Magnetic Field \cite{28}, and Finally the two-dimensional Dirac oscillator
in both commutative and non-commutative phase-space\cite{29,30}.

The principal aim of this paper is to study the effect of the presence
of a nonzero minimal length on the thermal properties of the Dirac
oscillator in one and two dimensions. For this, we use the formalism
based on the Epstein Zeta function to calculate the canonical partition
function in both cases. We expect that the introduction of a minimal
length have important consequences on these properties.

This paper is organized as follows: in sec. II, we propose a method
based on Epstein Zeta function to calculate the canonical partition
function of the Dirac oscillator in one and two dimensions. Sec. III
is devoted to present the different results concerning the thermodynamics
quantities of this oscillator. Finally, sec. VI will be a conclusion.

\section{Zeta Thermal partition function of a Dirac oscillator in one and
two dimensions}

\subsection{Framework theory}

The two-dimensional zeta Epstein function $\mathcal{Z}$ is defined
for $\mbox{\ensuremath{\mathcal{R}}e}\,s>1$, by \cite{31,32,33,34}
\begin{equation}
\mathcal{Z}\left(s\right)=\sum_{n,m=-\infty}^{\infty}\frac{1}{\left(am^{2}+bmn+cn^{2}\right)^{s}},\label{eq:1}
\end{equation}
where $a,\,b,\,c$ are real numbers with $a>0$ and $D=b^{2}-4ac$.
By defining that $D=4ac-b^{2}>0$, then the following quantity
\begin{equation}
Q\left(m,n\right)=am^{2}+bmn+cn^{2},\label{eq:2}
\end{equation}
is a positive-definite binary quadratic form of discriminant $D$.
In this case, we have
\begin{equation}
\mathcal{Z}\left(s\right)=\sum_{n,m=-\infty}^{\infty}\frac{1}{Q\left(m,n\right)^{s}}.\label{eq:3}
\end{equation}
Now, by using the following substitutions 
\begin{equation}
x=\frac{b}{2a},\,y=\frac{\sqrt{D}}{2a},\,\tau=x+iy,\label{eq:4}
\end{equation}
Eq. (\ref{eq:3}) 
\begin{equation}
\mathcal{Z}\left(s\right)=\sum_{n,m=-\infty}^{\infty}\frac{1}{a^{s}\mid m+n\tau\mid^{2s}}.\label{eq:5}
\end{equation}
By following the procedure used in \cite{31}, the final form of two-dimensional
Epstein zeta function is
\begin{equation}
\mathcal{Z}\left(s\right)=2a^{-s}\zeta\left(2s\right)+2a^{-s}y^{1-2s}\sqrt{\pi}\frac{\zeta\left(2s-1\right)\Gamma\left(s-\frac{1}{2}\right)}{\Gamma\left(s\right)}+\frac{2a^{-s}y^{\frac{1}{2}-s}\pi^{s}}{\Gamma\left(s\right)}H\left(s\right),\label{eq:6}
\end{equation}
with \cite{31}
\begin{equation}
H\left(s\right)=4\sum_{k=1}^{\infty}\sigma_{1-2s}\left(k\right)k^{s-\frac{1}{2}}\cos\left(2k\pi x\right)K_{s-\frac{1}{2}}\left(2k\pi y\right),\label{eq:7}
\end{equation}
where $\sigma_{\nu}\left(k\right)$ denotes the sum of the $v$-th
powers of the divisors of $k$, that is, 
\begin{equation}
\sigma_{\nu}\left(k\right)=\sum_{d/k}d^{\nu}=\sum_{d/k}\left(\frac{k}{d}\right)^{\nu}.\label{eq:8}
\end{equation}

\subsection{The zeta thermal function}

We start with the following eigenvalues of a one-dimensional Dirac
oscillator in the presence of minimal length $\beta$ \cite{21}
\begin{equation}
\epsilon_{n}=m_{0}c^{2}\sqrt{1+2\frac{\hbar\omega}{m_{0}c^{2}}n+\beta\frac{\hbar^{2}\omega^{2}}{c^{2}}n^{2}}.\label{eq:9}
\end{equation}
With the substitutions
\begin{equation}
b=2r,a=r^{2}\frac{\beta}{\beta_{0}},\,\left(r=\frac{\hbar\omega}{m_{0}c^{2}},\beta_{0}=\frac{1}{m_{0}^{2}c^{2}}\right),\label{eq:10}
\end{equation}
we get
\begin{equation}
\epsilon_{n}=m_{0}c^{2}\sqrt{an^{2}+bn+1}.\label{eq:11}
\end{equation}
In what follow, we choose $r=1$. Given the energy spectrum, we can
define the partition function via
\begin{equation}
Z_{1D}=\sum_{n}e^{-\tilde{\beta}\epsilon_{n}},\label{eq:12}
\end{equation}
where $\tilde{\beta}=\frac{1}{k_{B}T}$ with $k_{B}$ is the Boltzmann
constant. In our case, $Z$ reads
\begin{equation}
Z_{1D}=\sum_{n}e^{-\frac{1}{\tau}\sqrt{an^{2}+bn+1}}.\label{eq:13}
\end{equation}
with $\tau=\frac{k_{B}T}{m_{0}c^{2}}$. Now, we put that 
\begin{equation}
\chi=\frac{1}{\tau}\sqrt{an^{2}+bn+1},\label{eq:14}
\end{equation}
and by using the following relation\cite{35} 
\begin{equation}
e^{-\chi}=\frac{1}{2\pi i}\int_{C}ds\chi^{-s}\Gamma\left(s\right),\label{eq:15}
\end{equation}
the sum is transformed into
\begin{equation}
\sum_{n}e^{-\frac{1}{\tau}\sqrt{an^{2}+bn+1}}=\frac{1}{2\pi i}\int_{C}ds\left(\frac{1}{\tau}\right)^{-s}\sum_{n}\left\{ an^{2}+bn+1\right\} ^{-\frac{s}{2}}\Gamma\left(s\right)=\frac{1}{2\pi i}\int_{C}ds\left(\frac{1}{\tau}\right)^{-s}\mathcal{Z}\left(s\right)\Gamma\left(s\right),\label{eq:16}
\end{equation}
and $\Gamma\left(s\right)$ and $\mathcal{Z}\left(s\right)$ are respectively
the Euler and one-dimensional Epstein zeta function \cite{32}, where
\begin{equation}
\mathcal{Z}\left(s\right)=\sum_{n}\frac{1}{Q\left(1,n\right)^{\frac{s}{2}}},\label{eq:17}
\end{equation}
with
\begin{equation}
Q\left(1,n\right)=an^{2}+bn+1.\label{eq:18}
\end{equation}
Setting that 
\begin{equation}
x=\frac{b}{2},\,y=\frac{\sqrt{D}}{2},\,\mbox{with}\,D=4a-b^{2}>0,\label{eq:19}
\end{equation}
we find the restriction on the deformation parameter $\beta$ 
\begin{equation}
\beta>\beta_{0}=\frac{1}{m_{0}^{2}c^{2}},\label{eq:19.1}
\end{equation}
and consequently, (\ref{eq:17}) is transformed into 
\begin{equation}
\mathcal{Z}\left(s\right)=2a^{-\frac{s}{2}}\zeta\left(s\right)+\frac{2a^{-\frac{s}{2}}y^{1-s}\sqrt{\pi}}{\Gamma\left(\frac{s}{2}\right)}\zeta\left(s-1\right)\Gamma\left(\frac{s}{2}-\frac{1}{2}\right)+\frac{2a^{-\frac{s}{2}}y^{\frac{1}{2}-\frac{s}{2}}\pi^{\frac{s}{2}}}{\Gamma\left(\frac{s}{2}\right)}H\left(\frac{s}{2}\right).\label{eq:20}
\end{equation}
Now, the final partition function is
\begin{equation}
Z_{1D}=\frac{1}{2\pi i}\int_{C}ds\left(\frac{1}{\tau}\right)^{-s}\mathcal{Z}\left(s\right)\Gamma\left(s\right),\label{eq:21}
\end{equation}
or
\begin{eqnarray}
Z_{1D} & = & \frac{1}{2\pi i}\int_{C}ds\left(\frac{1}{\tau}\right)^{-s}2a^{-\frac{s}{2}}\zeta\left(s\right)\Gamma\left(s\right)+\frac{1}{2\pi i}\int_{C}ds\left(\frac{1}{\tau}\right)^{-s}\frac{2a^{-\frac{s}{2}}y^{1-s}\sqrt{\pi}}{\Gamma\left(\frac{s}{2}\right)}\zeta\left(s-1\right)\Gamma\left(\frac{s}{2}-\frac{1}{2}\right)\Gamma\left(s\right).\nonumber \\
 &  & +\frac{1}{2\pi i}\int_{C}ds\left(\frac{1}{\tau}\right)^{-s}\frac{2a^{-\frac{s}{2}}y^{\frac{1}{2}-\frac{s}{2}}\pi^{\frac{s}{2}}}{\Gamma\left(\frac{s}{2}\right)}H\left(\frac{s}{2}\right)\Gamma\left(s\right)\label{eq:22}
\end{eqnarray}
The first integral has two poles in $s=0$ and $s=1$, the second
has three poles in $s=0$, $s=1$ and $s=2$, and finally the third
has a pole at $s=0$. By Applying the residues theorem, we get
\begin{equation}
Z_{1D}=2\zeta\left(0\right)+\frac{2}{\sqrt{a}}\left\{ \zeta\left(1\right)+\zeta\left(0\right)\right\} \tau+\frac{2\pi}{ay}\tau^{2}.\label{eq:23}
\end{equation}
The last integral goes to the zero because of the following relations
\begin{equation}
\frac{1}{\Gamma\left(s\right)}=se^{\gamma s}\prod_{n=1}^{\infty}\left\{ \left(1+\frac{x}{n}\right)e^{-\frac{x}{n}}\right\} ,\label{eq:24}
\end{equation}
with where $\gamma$ is Euler 's constant given by
\begin{equation}
\gamma=\lim_{n\rightarrow\infty}\left(\sum_{k=1}^{n}\frac{1}{k}-\log\left(n\right)\right),\label{eq:25}
\end{equation}
Elizalde \cite{32,33} also mentioned that this formula is very useful
and its practical application quite simple: in fact, the two first
terms are just nice, while the last one is quickly convergent and
thus absolutely harmless in practice.

Thus,the final partition function for the one-dimensional Dirac oscillator
becomes
\begin{equation}
Z_{1D}\left(\tau,\alpha\right)=\frac{2\pi}{\alpha\sqrt{\alpha-1}}\tau^{2}+\frac{1}{\sqrt{\alpha}}\tau-1,\label{eq:26}
\end{equation}
with $\alpha=\frac{\beta}{\beta_{0}},$ so $a=\alpha$ and $y=\sqrt{\alpha-1}$.
We note here that the case of a two-dimensional can be treated in
the same way as that used in one dimension: starting with the following
form of the spectrum of energy (see \cite{29})
\begin{equation}
\bar{\epsilon}_{n}=m_{0}c^{2}\sqrt{1+4\frac{\hbar\omega}{m_{0}c^{2}}n+4\beta\frac{\hbar^{2}\omega^{2}}{c^{2}}n^{2}},\label{eq:27}
\end{equation}
and by the same procedure as described above, a final wanted partition
function of a two-dimensional Dirac oscillator is
\begin{equation}
Z_{2D}\left(\tau,\alpha\right)=\frac{\pi}{4\alpha\sqrt{\alpha-1}}\tau^{2}+\frac{1}{2\sqrt{\alpha}}\tau-1.\label{eq:28}
\end{equation}
Finally, all thermal properties for both cases can be obtained by
using the following relations
\begin{equation}
\mathcal{F}\equiv\frac{F}{mc^{2}}=-\tau\ln\left(Z\right),~\mathcal{U}\equiv\frac{U}{mc^{2}}=\tau^{2}\frac{\partial\ln\left(Z\right)}{\partial\tau},\label{eq:29}
\end{equation}
\begin{equation}
\mathcal{S}\equiv\frac{S}{k_{B}}=\ln\left(Z\right)+\tau\frac{\partial\ln\left(Z\right)}{\partial\tau},~\mathcal{C}\equiv\frac{C}{k_{B}}=2\tau\frac{\partial\ln\left(Z\right)}{\partial\tau}+\tau^{2}\frac{\partial^{2}\ln\left(Z\right)}{\partial\tau^{2}}.\label{eq:30}
\end{equation}

\section{Numerical results and discussions}

\begin{figure}[H]
\subfloat[1D Dirac oscillator]{\includegraphics[scale=0.45]{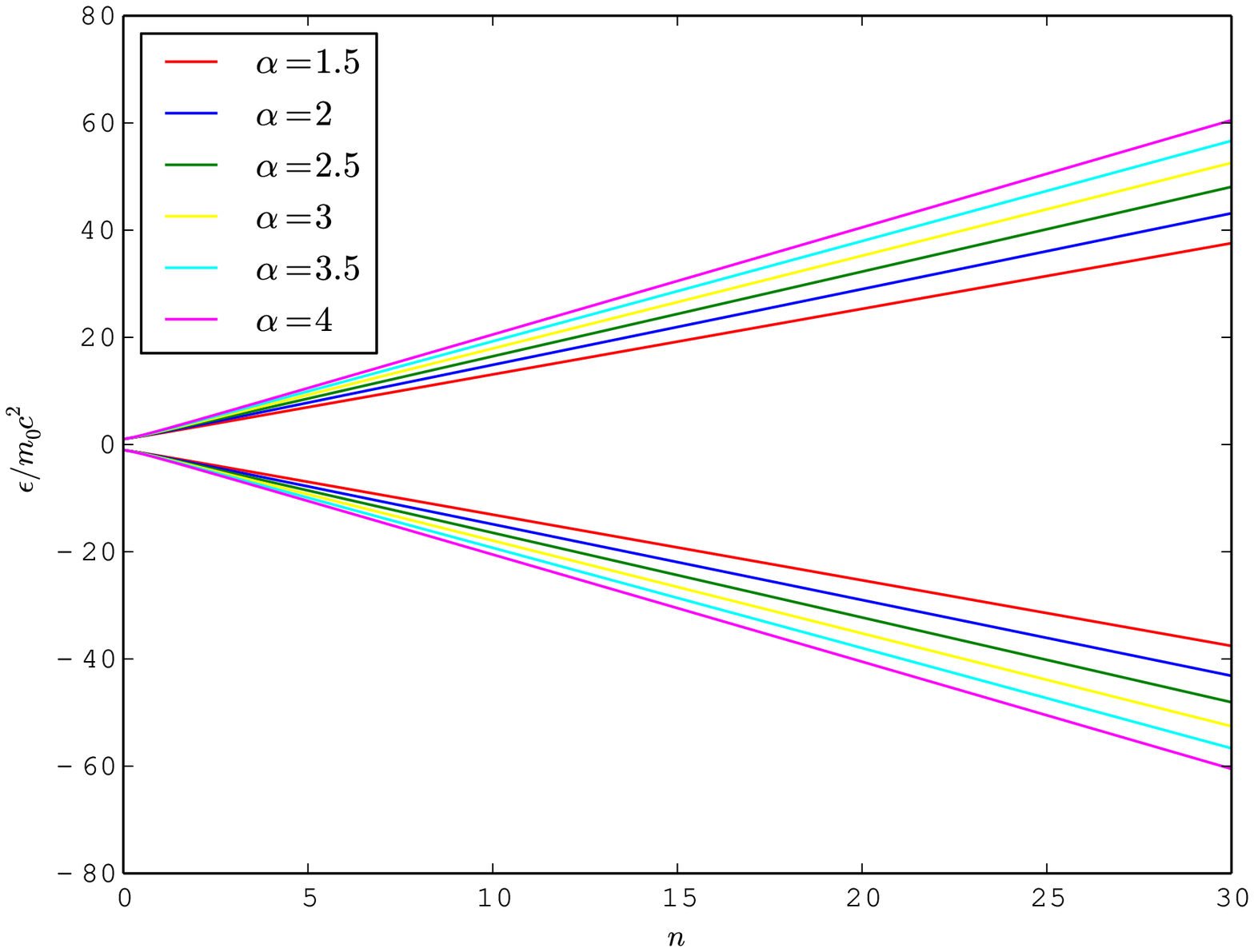}

}~\subfloat[2D Dirac oscillator]{\includegraphics[scale=0.45]{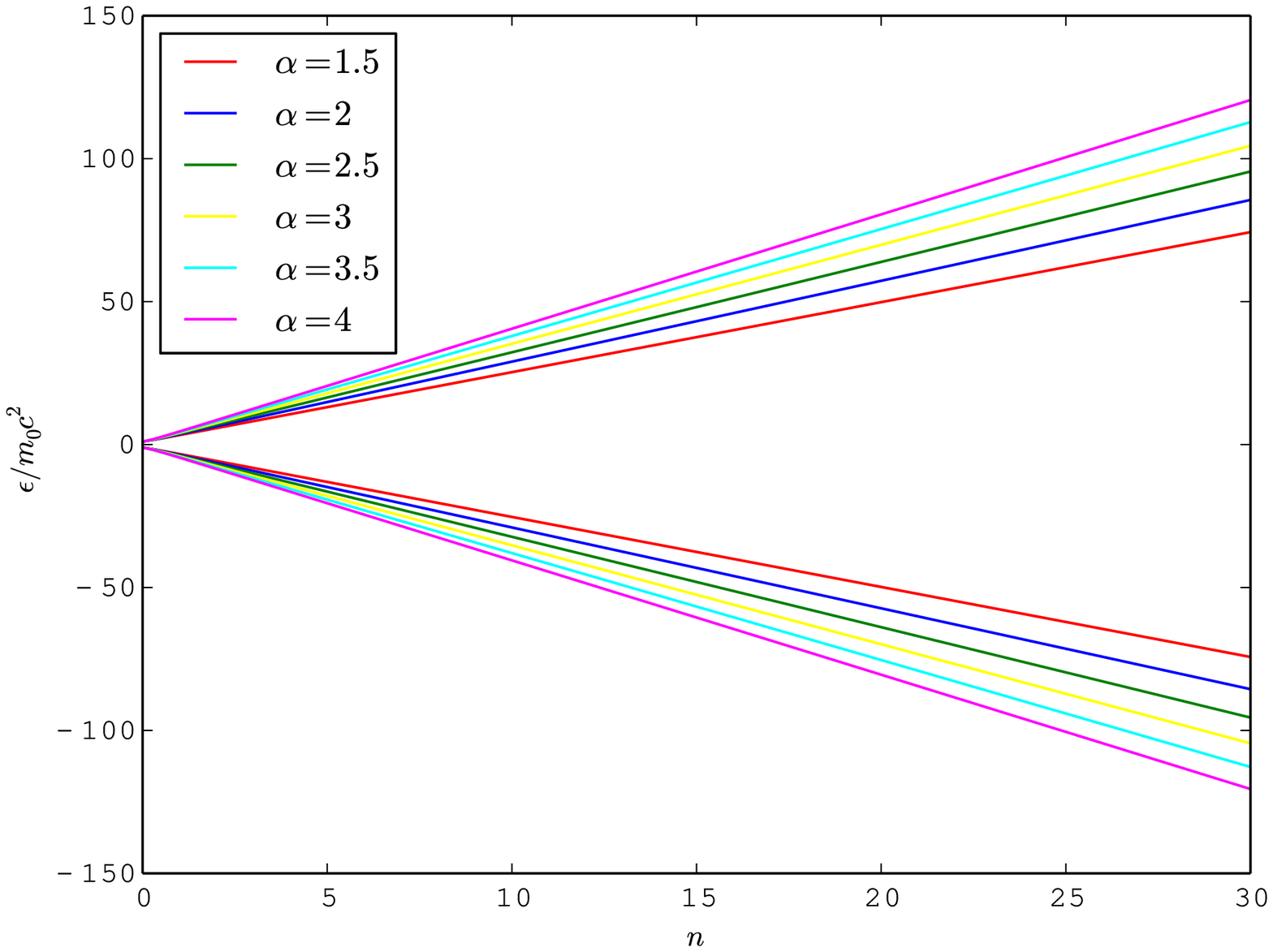}

}

\caption{\label{fig1}Energy spectrum $\frac{\epsilon}{m_{0}c^{2}}$ versus
quantum number $n$ for different values of $\alpha=\frac{\beta}{\beta_{0}}$.}
\end{figure}

Before presenting our results concerning the thermal quantities of
one and two dimensional Dirac oscillator, two remarks can be made:
(i) in Table. \ref{tab:Some-values-of}, we show some values of $\beta_{0}$
together with the minimal length $\Delta x$ for some fermionic particles:
\begin{table}[h]
\begin{tabular}{|c|c|c|c|c|}
\cline{2-5} 
\multicolumn{1}{c|}{} & Symbol & Mass$\left(\frac{\mbox{MeV}}{\mbox{c}^{2}}\right)$ & $\beta_{0}=\frac{1}{m_{0}^{2}c^{2}}\left(\mbox{\ensuremath{\frac{s^{2}}{kg^{2}m^{2}}}}\right)$ & $\Delta x\simeq\hbar\sqrt{\beta_{0}}\left(\mbox{m}\right)$\tabularnewline
\hline 
electron/positron & $e^{-}/e^{+}$ & $0.511$ & $1.339109486\times10^{43}$ & $38.615926800\times10^{-14}$\tabularnewline
\hline 
proton/anti-proton & $p/\bar{p}$ & $938.272$ & $3.971566887\times10^{36}$ & $2.1030891047\times10^{-16}$\tabularnewline
\hline 
muon & $\mu^{-}/\mu^{+}$ & $105.7$ & $3.143705046\times10^{38}$ & $1.867594294\times10^{-15}$\tabularnewline
\hline 
tauon & $\tau^{-}/\tau^{+}$ & $1777$ & $1.105704217\times10^{36}$ & $1.11056\times10^{-16}$\tabularnewline
\hline 
\end{tabular}

\caption{\label{tab:Some-values-of}Some values of both $\beta_{0}$ and minimal
length $\Delta x=\sqrt{\beta_{0}}$ .}
\end{table}
. This parameter has been determinate through the properties of Epstein
Zeta function, and this restriction leads to the minimal length $\Delta x\simeq\hbar\sqrt{\beta_{0}}$
. According to Table. \ref{tab:Some-values-of} the minimal length
lies in the interval $10^{-16}<\Delta x<10^{-14}\mbox{m}$. In addition,
we can see that $\Delta x\simeq\hbar\sqrt{\beta_{0}}=\frac{\hbar}{m_{0}c}=\bar{\lambda}$,
where $\bar{\lambda}$ is the reduced Compton wavelength: so the minimal
length has the same order as the reduced Compton wavelength. (ii)
in Figure. \ref{fig1} we study, the effect of the presence of minimal
length on the spectrum of energy. In this context, the reduced spectrum
of energy as a function of the quantum number $n$ for different values
of $\alpha$ are depicted in fig. \ref{fig1}. This figure reveals
that the effect deformation parameter $\beta$ on the energy spectrum
is significant.

Moreover, we note that we have only restrict ourselves to stationary
states of positive energy. The reason for this is twofold \cite{36}:
(i) the Dirac oscillator possesses an exact Foldy\textendash Wouthuysen
transformation (FWT): so, the positive- and negative-energy solutions
never mix. (ii) The solutions with infinite degeneracy do not correspond
to physical states since there is not Lorentz finite representation
for them. Thus, according to these arguments, we can assume that only
particles with positive energy are available in order to determine
the thermodynamic properties of our oscillator in question.

\begin{figure}[H]
\includegraphics[scale=0.8]{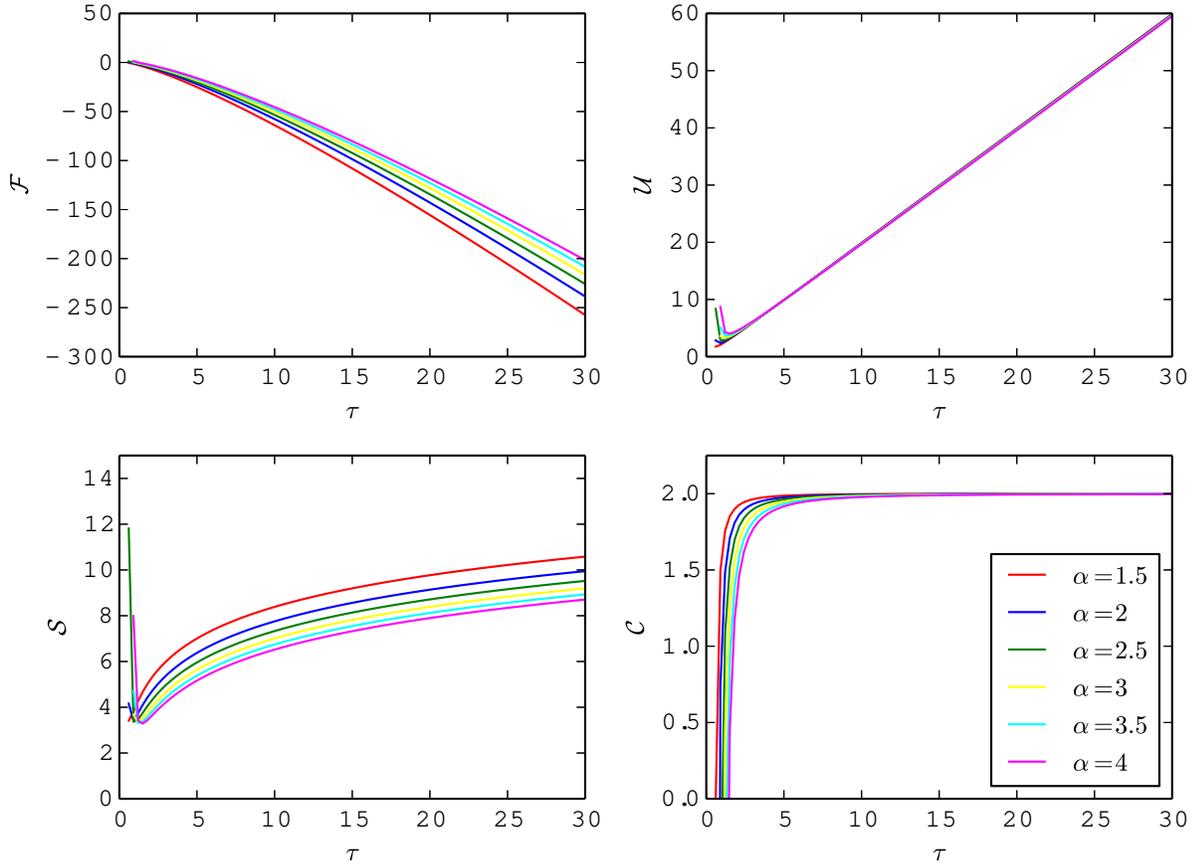}

\caption{\label{fig2}Thermal properties of a one dimensional Dirac oscillator
for different values of $\alpha=\frac{\beta}{\beta_{0}}$.}
\end{figure}

Now, we are ready the present our numerical results on the thermal
properties of a Dirac oscillator in one and two dimensions: in Fig.
\ref{fig2}, we show all thermal properties of the one dimensional
Dirac oscillator for different values of $\alpha$. According to this
figure, we can confirm that the parameter $\beta$ plays a significant
role on these properties, and the effect of this parameter is very
important on the thermodynamic properties. In particularity, the curves
of the reduced specific heat, for different values of $\beta$, tend
to the an asymptotic limit at $2$ , and they separated in the range
of the reduce temperature $\tau$ between $0$ and $\sim10$ .

\begin{figure}[H]
\includegraphics[scale=0.8]{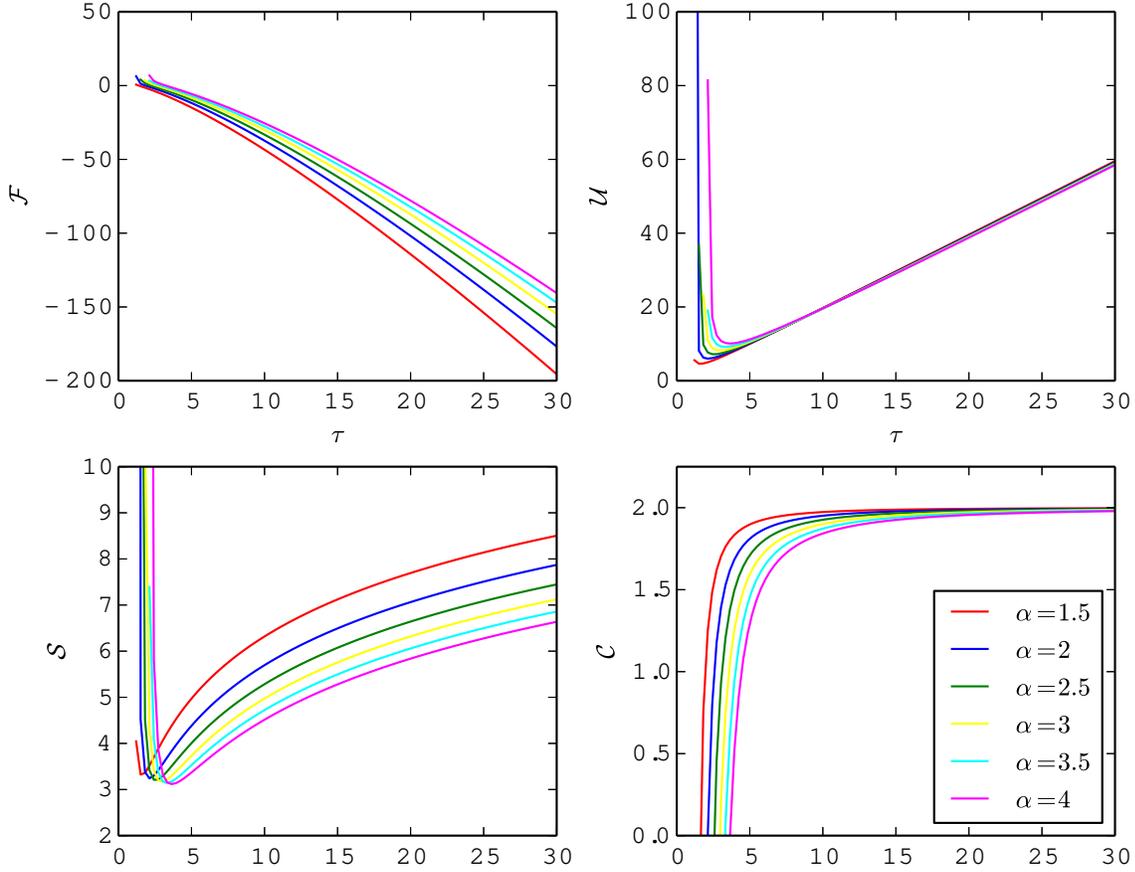}

\caption{\label{fig3}Thermal properties of a two dimensional Dirac oscillator
for different values of $\alpha=\frac{\beta}{\beta_{0}}$.}
\end{figure}

For the case of a two-dimensional oscillator, and according to the
Eqs. (\ref{eq:9}) and (\ref{eq:27}), we conclude that the method
of determining the canonical partition function will be the same in
both cases. As a consequence, all thermal properties can be found
by the same manner as in the one-dimensional case. These properties
are depicted in Fig. \ref{fig3}
, and can be explained by the same way described in the one-dimensional
case.

\section{Conclusion}

In this work, we have study the influence of the minimal length on
the thermal properties of a Dirac oscillator in one and two dimensions.
The statistical quantities of both cases were investigated by employing
the Zeta Epstein function method. All this properties such as the
free energy, the total energy, the entropy, and the specific heat,
show the important effect of the presence of minimal length on the
thermodynamics properties of a Dirac oscillator. Moreover, the formalism
based on the properties of Zeta Epstein function allows us to calculate
the values of minimal length $\triangle x=\hbar\sqrt{\beta}$ for
some fermionic particles as shown in Table. \ref{tab:Some-values-of}.
These values coincide well with the reduced Compton wavelength $\bar{\lambda}$. 
\begin{acknowledgments}
The authors wish to thank Prof Emilio Elizalde for his helpful comments
and discussions about the Epstein Zeta Function.\end{acknowledgments}

\end{document}